%% file: ms.tex
\documentclass{emulateapj}

\begin{document}

\newcommand{\kms}{\,km\,s$^{-1}$}
\newcommand{\vej}{v_{\mathrm{ej}}}
\newcommand{\vmax}{v_{\mathrm{max}}}
\newcommand{\acd}{N_{\mathrm{a}}}
\newcommand{\zabs}{z_{\mathrm{abs}}}
\newcommand{\zem}{z_{\mathrm{em}}}

\slugcomment{Accepted for Publication in {\it The Astrophysical
Journal}}

\title{On the Fraction of Quasars with Outflows}
\author{Rajib Ganguly \& Michael S. Brotherton\altaffilmark{1}}
\altaffiltext{1}{Department of Physics \& Astronomy, The University
of Wyoming (Dept. 3905), 1000 East University Ave., Laramie, WY,
82071}
\begin{abstract}
Outflows from active galactic nuclei (AGNs) seem to be common and
are thought to be important from a variety of perspectives: as an
agent of chemical enhancement of the interstellar and intergalactic
media, as an agent of angular momentum removal from the accreting
central engine, and as an agent limiting star formation in
starbursting systems by blowing out gas and dust from the host
galaxy.  To understand these processes, we must determine what
fraction of AGNs feature outflows and understand what forms they
take.  We examine recent surveys of quasar absorption lines,
reviewing the best means to determine if systems are intrinsic and
result from outflowing material, and the limitations of approaches
taken to date. The surveys reveal that, while the fraction of
specific forms of outflows depends on AGN properties, the overall
fraction displaying outflows is fairly constant, approximately
60$\%$, over many orders of magnitude in luminosity.  We emphasize
some issues concerning classification of outflows driven by data
type rather than necessarily the physical nature of outflows, and
illustrate how understanding outflows probably requires more a
comprehensive approach than has usually been taken in the past.
\end{abstract}


\section{Introduction}

The role of outflows from quasars and active galactic nuclei (AGN)
has recently become an important feature in the overall framework of
how galaxies and star formation processes evolve over cosmic time.
Mergers and other interactions triggering AGN seem to provide
feedback affecting the larger scale environment. Recent efforts to
include the effects of this so-called AGN feedback focus on two
modes: a ``radio'' mode whereby a relativistic jet heats the
surrounding interstellar and intercluster media
\citep[e.g.,][]{best07}, and a ``quasar'' mode whereby a lower
velocity but higher mass outflow also helps to clear out post-merger
shrouding gas and quenches star formation \citep*[e.g.,][]{tdm05}.
We focus on this second mode in this paper. For this mode, a number
of questions require addressing. How common are outflows? Do all AGN
have outflows? What drives outflows? Is there a single all-governing
structure of AGN? Answering these questions will help us to
understand the role AGN outflows with respect to issue of feedback,
and other important issues like chemical enrichment and accretion.

In the ensuing sections we aim to achieve several goals: (1) to
review the ways in which outflows are detected in AGN over all
luminosity scales; (2) to comment on the merits of various catalogs
of outflows; and (3) to arrive at the true (possibly
property-dependent) observed frequency of outflows. In its most
basic interpretation, the observed frequency of outflows can be
equated with the fraction of solid angle (from the view point of the
central black hole) subtended by outflowing gas. This interpretation
assumes that all AGN feature outflows and that not all sight-lines
to the emitting regions are occulted by the outflow. Alternatively
(and equally simplistic), the frequency can be interpreted as the
fraction of the duty cycle over which AGN feature outflows (assuming
the outflow subtends 4$\pi$\ steradians). The actual conversion of
the fraction of AGN featuring spectroscopic evidence of outflow to
the solid angle subtend by such outflows has been treated by
\citet{cren99} and \citet*{cren03b}. This computation involves
further knowledge of the line-of-sight covering factor (that is, the
fraction of lines-of-sight that reach the observer that are occulted
by the outflow) as well as an understanding of range of solid angle
sampled by the AGN used (e.g., Type 1 versus Type 2 AGN). The true
situation is likely in between these two extremes, and may depend
also on properties we can not currently constrain, such as the time
since the AGN was triggered.

Additionally, we strive here to build a case that more effort should
be made to consider outflows of all types together.  Often data
limitations of one sort or another have led to the study of limited
parts of parameter space (e.g., outflow velocity or velocity
dispersion), creating artificial or at least biased divisions.  There
appears to be a continuous range in properties of outflows and these
should only be regarded as fundamentally different when there is clear
evidence to reach such a conclusion.  Below we discuss the
identification of outflows (\S 2) and the data-driven subcategories
(\S 3).  We show an illustrative example of how combining the
different outflow subclasses may lead to a more unified physical
understanding of outflows (\S 4).  Finally, we bring together the
different survey methodologies to determine an overall fraction of AGN
displaying the signatures of outflows (\S 5) and summarize the case
for more global studies of the outflow phenomenon.  We adopt a
cosmology with $\Omega_M$ = 0.3, $\Omega_{\Lambda}$ = 0.7, and H$_0$ =
70 km s$^{-1}$ Mpc$^{-1}$.

\section{The Ideal Way To Select Quasars Exhibiting Outflows}

Outflows from AGN are primarily detected in ultraviolet and X-ray
absorption against the compact continuum source (i.e, the inner
portions of the accretion disk) and/or the more extended broad
emission line region. In a few cases, outflows have been
demonstratively observed in emission both from the broad line region
in narrow-line Seyfert 1 galaxies
\citep[e.g.,][]{leighly04a,leighly04b,yuan07} and from the narrow
line region of Seyfert 1 galaxies \citep[e.g.,][]{das05,das06}.
[Arguably, the fact that broad emission lines in most AGN have only
a single peak is also a signature of outflowing gas
\citep[e.g.,][]{mur95}.] For emission-line gas, reverberation
mapping provides a direct means at establishing the location of the
gas. For absorption-line gas, placing a distance between the gas and
the ionizing continuum relies on using absorption-line diagnostics
to assess the photoionization parameter ($U$), and having other
information that constrains the density ($n$) of the gas. The
distance, $r$, is related to these quantities via
\begin{equation}
r = \sqrt{{{\int_{\nu_{\mathrm{LL}}}^{\infty} (L_\nu/\nu)  d\nu}
\over {4 \pi h c n U}}},
\end{equation}
where $\nu_\mathrm{LL} = 3.3 \times 10^{15}$\,Hz is the frequency of
the Lyman limit. Constraints on the density can come from
time-variability (if ionization/recombination dominates the
variability timescale), or the presence of excited-state lines.
Density information is not typically available for intrinsic
absorbers, so secondary indicators must be employed to separate
intrinsic absorbers from absorption by interloping structures (e.g.,
IGM filaments, galaxy halos and disks).

In order of decreasing utility and importance, the secondary
indicators of an intrinsic origin for an absorption-line system are:
(1) velocity width, (2) partial coverage, (3) time variability,  (4)
high photoionization parameter, (5) high metallicity,
\citep[e.g.][]{bs97}. Not all intrinsic absorbers exhibit all of
these properties, but the probability of an intrinsic origin is
higher if an absorber exhibits more than one property. Likewise,
with the exception of the first two indicators (and the first only
in its most extreme, see \S\ref{sec:bal}), each of these indicators
have been observed in intervening material. Thus, by themselves, no
one indicator should be taken to imply an intrinsic origin.

Historically, the first criterion has led to three divisions in the
classification of intrinsic absorbers. Outflows with the largest
velocity dispersions are termed ``broad absorption lines''
\citep[e.g.,][BALs, FWHM $\geq2000$\,\kms]{weymann91}. On the other
extreme, intrinsic absorbers where the velocity dispersion is
sufficiently small as to cleanly separate the \ion{C}{4} doublet are
called ``narrow absorption lines'' \citep[e.g.,][NALs, FWHM
$\leq500$\,\kms]{hf99}. Since there is a whole continuum of velocity
widths, this has led to an in-between class known as ``mini-BALs''
\citep[e.g.,][]{ham97c,cssg99}. Below we examine each of these
classes in terms of their observed frequency, and note various
issues in determining this number, including dependencies on quasar
physical properties.

\section{Observations of the Incidence of Various Forms of Outflows}

\subsection{Broad Absorption Lines (BALs)}
\label{sec:bal}

Broad absorption lines in the spectra of quasars are the most easily
identifiable forms of outflows. The large velocity width is very
readily associated with accelerated, outflowing gas. As such, these
garnered more attention historically than their smaller
velocity-width kin
\citep[e.g.][]{weymann85,turn88,turn88b,weymann91,vwk93}.

\citet{weymann91} established criteria, summarized in a number
called the BALnicity index (BI), for determining if an absorption
line constituted a BAL. The BI was a modified form of an equivalent
width whereby one counted absorption that fell below 90\% of the
true quasar continuum that was contiguously below this level for
more than 2000\,\kms. Moreover, no absorption within 3000\,\kms\ of
the quasar redshift was counted in order to remove possible
contamination by absorption-line gas not physically associated with
the quasar central engine (e.g., interstellar gas from the quasar
host galaxy, or intergalactic material from the host cluster).
[Note: With a minimum velocity of 3000\,\kms and a minimum
contiguous width of 2000\,\kms, this means that no absorption
falling entirely within 5000\,\kms\ of the quasar redshift is
counted.] This index was established using low-dispersion data of
high-redshift ($1.5 \leq z \leq 3.0$) objects from the Large Bright
Quasar Survey, or LBQS \citep{lbqs1,foltz89,lbqs3,lbqs6}, and was designed to
yield a pure sample of objects with bonafide outflows. We note here
that the utility of BI was driven purely by the data quality (signal-to-noise
ratio and resolution) of the LBQS spectra in conjunction with the desire
to remove false positives (at the expense of losing some true BAL quasars).
While the use of BI to define samples of BAL quasars has utility, especially
in comparing results between data sets of varying quality, it
excludes some fraction of real high-velocity dispersion outflows that
qualitatively appear to be BAL quasars but just fail to have positive BI.

An improvement on the BI, termed the intrinsic absorption index (AI),
was developed by \citet{hallai} to alleviate the inadequacies of BI in
selecting objects where high velocity outflows were clearly observed
but were not included as BAL quasars by the BI criteria (e.g., UM 660,
PG 2302+029). The AI was designed to be more flexible and inclusive
and has been very useful in its application to newer and better
quality datasets like the Sloan Digital Sky Survey (SDSS). This
flexibility, while good at including objects not previously selected
by BI, has increased the contamination of samples of intrinsic
absorption while still not including other forms of intrinsic
absorption \citep[e.g.,][]{gan07b}.

The incidence of BALs has primarily been determined using optical
spectra where, historically, large samples of high-redshift quasars
(to get rest-frame UV coverage) could efficiently be selected (e.g.,
with color-selection). In such surveys
\citep{hf03,reichard03,trump06,gan07b}, roughly 10-25\% of objects are
observed to host BALs. An issue with optical/UV surveys, however, is
potential biases in the selection of quasars against those hosting
BALs due to the fact that much of the continuum is absorbed
\citep[e.g.,][]{good95,goodrich97,kv98} and intrinsically reddened
\citep[e.g.,][]{reichard03}.  Using the LBQS, where the observed
frequency of BAL quasars in the redshift range $1.5 \leq z \leq 3$\ is
15\% using a BI criterion, \citet{hf03} estimated a true BAL frequency
of 22\% from comparisons in the $k$-corrections of BAL and non-BAL
quasars. The recent catalog of BAL quasars using an AI criterion from
\citet{trump06} found a BAL frequency of 26\% (in the redshift range
$1.7 \leq z \leq 4.38$). Both of these estimates are based on the
\ion{C}{4}\ $\lambda1548,1550$\ doublet, which is the most commonly
used species in selecting intrinsic absorption owing to the relatively
high abundance of carbon, the high ionization fraction of C$^{3+}$\ in
moderately-ionized gas, and the resonant absorption of the doublet.

To combat possible selection biases in the optical, one can examine
quasar catalogs selected in other bands. \citet{becker00} examined
radio-selected quasars from the FIRST Bright Quasar Survey and found
a BAL quasar frequency of about 18\% (though it is only 14\% if only
BI$>0$\ objects are counted, comparable to other estimates based on
optical-selection). This again predominantly used the \ion{C}{4}
doublet and objects at $z > 1.7$.

Incidentally, several studies \citep[e.g.,][]{bro98} have now
dismantled the myth that BALs are only observed in formally
radio-quiet (i.e.,
$f_\nu(\mathrm{5\,GHz})/f_\nu(\mathrm{3000\,\AA})<10$) objects, though
their frequency does significantly decrease among the most radio-loud
quasars \citep*{becker01,gregg06}. We note that a subset of
radio-selected BAL quasars can be identified as polar outflows
\citep*[e.g.,][]{zhou06,brotherton06,gp07}. At least one of these
objects, FIRST J155633.8$+$351758, appears to be an optically reddened
and beamed radio-quiet quasar \citep{rb07}.  The presence of BAL
outflows in such objects as well in as edge-on FR II BAL quasars,
\citep[e.g., ][]{gregg06} indicates high-velocity outflows are present
in a variety of geometries.  There is as yet no observational
signature in the absorption spectra that is correlated with
orientation indicators, so any geometrically restrictive model such as
those identifying BAL outflows solely with equatorial winds are either
wrong or incomplete.  Any complete picture of outflows must reflect a
range of geometries.  It has yet to be established observationally how
often polar outflows occur compared to equatorial, or if the location
or dynamics differ.

In addition to radio-selection, one can examine the frequency of BAL
quasars from infrared selection. Recently, \citet{dai07} compared the
catalog of BAL quasars \citep{trump06} from the Third Data Release
(DR3) of SDSS and the parent sample of DR3 quasars \citep{sdssqso3}
with the Two Micron All-Sky Survey \citep[][2MASS]{2mass} Point-Source
Catalog (PSC). With some variation with redshift, they reported an
overall true BAL quasar fraction of $43\pm2$\%, markedly higher than
estimates based on UV/optical data alone. Presumably, this difference
accounts for the effects of dust and absorption that may bias
UV/optical selection techniques against find BAL quasars.

We point out that this estimate relies heavily on the automated
techniques employed in finding BAL quasars in a large dataset such
as SDSS. From a critical look at 5088 $1.7 < z < 2$\ quasars from
SDSS DR2, \citet{gan07b} noted several instances of false (and
missed) classifications in the \citet{trump06} catalog. A comparison
of the \citet{gan07b} sample with the 2MASS PSC reveals a BAL
fraction of 66/287 (23\%), completely consistent with the analysis
of \citet{hf03}. Blindly using the \citet{trump06} catalog yields a
BAL fraction of 96/287 (33\%), consistent with the $z<2$\ points
from \citet[][see their Figure 4]{dai07}. At face value, this
implies that nearly 30\% of the \citet{trump06}-2MASS cross-matched
sample consists of false-positives. We return to the issue of
false-negatives below.



\subsection{Narrow Absorption Lines (NALs) and mini-BALs}

Intrinsic NALs and mini-BALs have, within the last decade, come to
light as a very powerful and complementary means of studying
outflows. Unlike their very broad kin, these absorbers are generally
not blended and, therefore, offer a means to determine ionization
levels and metalicities using absorption-line diagnostics. Thus,
NALs and mini-BALs are more useful as probes of the physical
conditions of outflows. The drawback, however, is that truly
intrinsic NALs and mini-BALs are more difficult to identify, since
interloping structures such as the cosmic web, galaxy clusters, and
galactic disks and halos also have comparable velocity spreads
($\lesssim800$\,\kms). Historically, progress was made by
statistically identifying an excess of absorbers over what is
expected from randomly distributed intervening structures
\citep[e.g.,][]{wwpt}. With improved technologies (such as
high-resolution spectroscopy with large telescopes), we can now take
advantage of the other secondary indicators to separate intrinsic
from intervening absorption. In the following subsections, we
discuss the frequency of two subclasses based on both historical and
more recent studies. We distinguish between absorbers that appear
near the quasar redshift (associated absorbers), and those that
appear at large velocity separations.

\subsubsection{Associated ($\zabs\sim\zem$) Absorbers (AALs)}
\label{sec:aal}

The term ``associated'' refers to narrow velocity-dispersion
absorption-line systems that lie near the quasar redshift. It has
been shown that the frequency of such systems is much larger than
those at large velocity separations
\citep{wwpt,foltz87,and87,ald94,rich99,rich01b}. Typically,
associated absorbers are defined as those lying within 5000\,\kms\
of the quasar redshift \citep{foltz86}. As such, they were
historically very complementary to BAL quasars selected using BI.
Updating BAL classification to reflect the better data quality
usually available today does allow for some confusion among classes,
at least in some cases, and this should be kept in mind. The issue
of what types of quasars hosted AALs was the subject of much
scrutiny with some studies claiming to see an excess of AALs
\citep[e.g.,][]{foltz87}, while other studies claimed no excess
\citep*[e.g.,][]{sbs88}. It was surmised that strong AALs (i.e.,
those with a large \ion{C}{4} equivalent width) were preferentially
found in optically-faint, steep radio spectrum quasars
\citep{mj87,foltz88}. However, more recent studies have found that
AALs are found (with varying frequency) in all AGN subclasses from
Seyfert galaxies \citep[e.g.,][]{cren99,kriss06} to higher luminosity
quasars \citep[e.g.][]{gan01a,lb02,ves03,misawa07}, and from steep
to flat radio spectrum sources \citep{gan01a,ves03}.

An important issue in the consideration of AALs as it relates to
outflows is where the absorbing gas originates. We note here a few
arguments for a direct association with outflows from the central
engine. While detailed studies of individual objects have shown
absorption-line {\it components} that must reside in the host galaxy
far from the central engine \citep[e.g.,][]{ham01,scott04,gan06a},
on the whole there have been no documented cases of AALs that are
truly redshifted with respect to the actual systemic velocity. If
AALs were to originate in the host galaxy, one would expect some
fraction of the absorbers to arise from infalling material. In fact,
the velocity distribution of \ion{C}{4} AALs is sharply peaked with
the \ion{C}{4} emission redshift \citep{gan01a}, implying a close
dynamical connection between AALs and the broad emission-line
region. In addition, blind studies of AALs using secondary
indicators find that $\geq20$\% are time-variable \citep{wise04},
and that $\sim33$\% show partial coverage \citep{misawa07}.

From an analysis of 59 $z<1$\ quasars, \citet{gan01a} showed that
the overall frequency of AALs was 25$\pm$6\%, with some variation
with broad-band spectral properties. Similar frequencies have been
established at higher redshift by \citet[][$27\pm5$\%]{ves03} and
\citet[][23\%]{misawa07}, both of which made attempts to filter out
contamination by intervening absorbers. Oddly, these fractions are
lower than the recent study of \citet{gan07b}, who find an AAL
frequency of 1898/5088 (37\%), although the 5000\,\kms\ velocity
cutoff for traditional AALs was not strictly adhered to in that
survey. We note that 1478/1898 (78\%) AALs in that study were missed
by the AI selection used by \citet{trump06}. These certainly
constitute false-negatives from the standpoint of finding intrinsic
absorption, though not from the standpoint of finding only BAL
quasars. While \citet{ves03} did note that quasars with AALs are
redder on average, a comparison of the \citet{gan07b} sample with
the 2MASS PSC reveals that the frequency of AALs is similar to the
parent sample (107/287, 37\%). Thus, the selection of AALs quasars
is not affected by optical biases (e.g., reddening or large optical
absorption) like BAL quasars.

\subsubsection{High Velocity NALs}

The first observational evidence for intrinsic narrow
velocity-dispersion absorption appearing at high ejection velocity
(many tens of thousands of kilometers per second) came nearly a
decade ago and include: PG\,2302+029 \citep{kp96}, Q\,2343+125
\citep*{ham97b}, and PG\,0935+417 \citep{ham97c}. Models of quasar
winds generally are able to explain outflows with $\Delta v/v \sim
1$, but are challenged by these $\Delta v/v << 1$\ systems. One idea
is that the sight-line cuts across the outflow that would produce a
BAL under a difference orientation \citep[e.g.,][]{elvis00,gan01a},
but this has yet to be demonstrated theoretically. These systems are
also interesting because they only absorb photons from the compact
continuum. Thus, partial coverage indicators provide severe
constraints on the geometry of the flow.

In terms of demographics, the first assessment of the frequency of
these systems came from \citet{rich99} and \citet{rich01b}. From a
statistical analysis examining the variation in the velocity
distribution of \ion{C}{4} NALs with quasar radio-loudness and
spectral index, \citet{rich99} estimated that as many as 36\% of
\ion{C}{4} NALs may arise from outflowing gas. Recently,
\citet{misawa07} report that only 10-17\% of \ion{C}{4} NALs in the
velocity range 5000-70000\,\kms\ show evidence of partial coverage.
(Thus it is possible that \citet{rich99} overestimated the fraction
of high-velocity NALs, or that 50-70\% of intrinsic \ion{C}{4} NALs
do not show partial coverage.) This is not a statement, however, on
the fraction of quasars that host such outflows.

\citet{ves03} reported that high velocity intrinsic NALs appeared in
18$\pm$4\% of $1.5 < z < 3.6$\ quasars in the velocity range
5000--21000\,\kms, with about a factor of two variation between
radio core-dominated (17$\pm$10\%) and radio lobe-dominated
(33$\pm$15\%) morphologies. In a recent survey of $1.8 < z < 3.5$\
SDSS sources, \citet{prh07} find about 12\% of quasars have
high-velocity NALs in the velocity range 5000-50000\,\kms, and
$\sim2.3$\% in the velocity range 25000-50000\,\kms. This latter
velocity range is often missed by surveys due purely to
observational cutoffs. Over this velocity range, absorption by
\ion{C}{4} can become confused with \ion{Si}{4} absorption.

\section{An Example Illustrating the Merits of Comprehensive Outflow Studies}

\begin{figure}
\epsscale{0.7} \rotatebox{-90}{\plotone{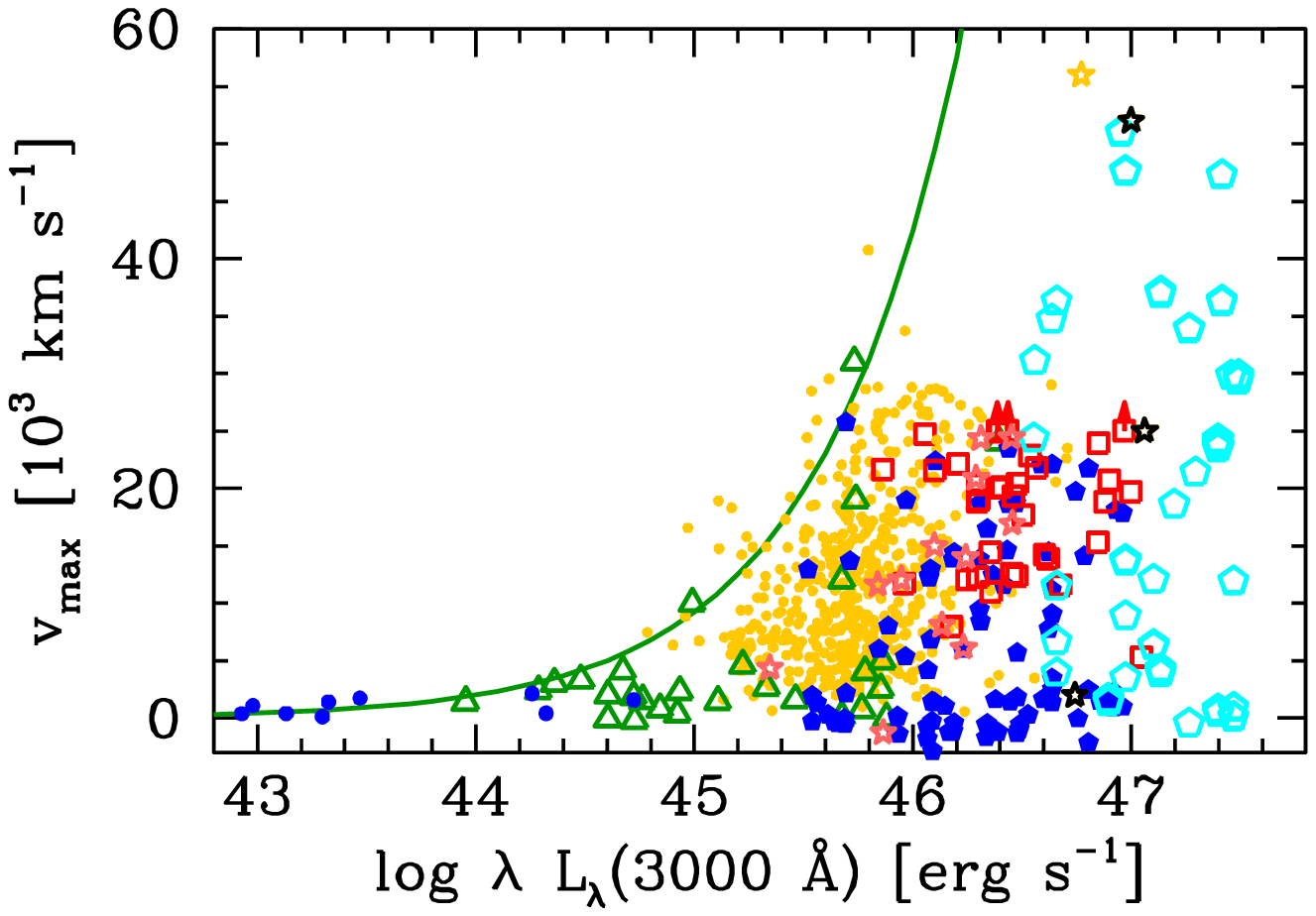}}
\protect\caption[vmax vs. luminosity]{We present a summary plot of
the maximum velocity of absorption versus the monochromatic
luminosity at 3000\,\AA. Data from several studies are included:
Seyfert 1 galaxies from \citet[][{\it filled blue circles}]{cren99};
$z<0.5$\ Palomar-Green quasars from \citet[][{\it open green
triangles}]{lb02}; SDSS DR2 BAL quasars from \citet[][{\it filled
yellow circles}]{gan07b}; LBQS BAL quasars from Gallagher et al.
(2006, {\it open red squares}); intrinsic NALs from \citet[][{\it
filled blue pentagons}]{ves03}; intrinsic NALs from \citet[][{\it
open cyan pentagons}]{misawa07}; polar BAL quasars from
\citet[][{\it pink stars}]{gp07}; UM 675, intrinsic NAL/mini-BALs in
Q\,2343+125, and PG\,0935+417 from \citet[][{\it black
stars}]{ham97c}; and the mini-BAL in PG\,2302+029 from \citet[][{\it
yellow star}]{kp96}.} \label{fig:vmaxlum}
\end{figure}

When parameter space is truncated, either intentionally (e.g.,
through subclass segregation or the desire to avoid false
positives/negatives) or unintentionally (e.g., by data limitations),
real correlations that could lead to physical understanding may be
missed.  The wide variety of observational techniques and the
improvements in sample size now make it possible to study the
outflow phenomenon in a more complete manner than ever before
possible.

\input{tab1.tex}

As discussed above, outflow velocity (both in terms of dispersion
and range) has often been limited in many studies.  By putting
together different surveys, we can study the velocity properties of
outflows as a function of other potentially important parameters.
One recent example where this has been shown to be fruitful is in
plotting maximum velocity of absorption, $\vmax$, against
luminosity.  This was originally done by \citet{lb02} for PG
quasars, who, finding an upper envelope formed by the soft X-ray
weak absorbers, argued for radiation-driven winds as quasar
outflows. More recently, others have added other samples of outflows
to this plot \citep[e.g.,][]{gallsc06,misawa07,gan07b} showing that
this envelope can be extrapolated to higher luminosities and
similarly constrains luminous BAL quasars.

We have further added to this plot in Figure~\ref{fig:vmaxlum}
\citep[with the envelope fit from ][]{gan07b}, including all types
of intrinsic outflows and the most extreme in terms of both terminal
velocity and luminosity. Polar BAL quasars are included as well and
fall among the general BAL quasar population. Intrinsic NAL systems
are well represented throughout the full luminosity range.  Some
data biases are present; for instance, the wavelength range of SDSS
spectra limited the observed $\vmax$\ to less than 30000 \kms\ for
most quasars studied by \citet{gan07b}, so the gap under the
envelope for luminosities above $10^{46}$\,erg\,s$^{-1}$\ is
potentially not real. Similarly, searching for intrinsic absorption
at velocities above 50000\,\kms\ becomes very difficult since the
\ion{C}{4} lines become blueshifted into the \ion{N}{5}/Lyman
$\alpha$ region where identification is especially challenging. Of
course, the empirical fit does not take into account relativistic
effects, so it is inappropriate to extrapolate it to arbitrarily
high luminosities. Taken at face value, the fit implies that a
quasar with luminosity $\lambda
L_\lambda$(3000\,\AA)$=10^{47.3}$\,erg~s$^{-1}$\ would be capable of
driving an outflow at the speed of light, but this is clearly
unphysical. More data, and insights establishing better criteria as
to which outflows sample the envelope are needed to improve our
understanding of the dependence on the terminal velocity on quasar
physical properties.

The figure also illustrates another (though more subtle) issue.
While AALs are typically defined as those NALs appearing within
5000\,\kms, some studies at lower redshift have opted for smaller
velocity differences, claiming that 5000\,\kms\ is an arbitrary
cut-off. The figure shows that there is a physical reason for this.
As the figure apparently shows, no outflow appears to be driven to a
velocity larger than is allowed by radiation pressure. At lower
redshifts, most objects studied are also lower luminosity (e.g.,
Seyfert galaxies). The figure clearly shows that objects with
$\lambda L_\lambda(3000\,\mathrm{\AA})\lesssim 3 \times
10^{44}$\,erg\,s$^{-1}$ are not capable of radiatively driving
outflows with velocities larger than 5000\,\kms. An insight such as
this is not only interesting for understanding AGN outflow physics,
but is also of use to other fields that make use of intervening
absorption-line systems (as it presents an additional means at
separating intrinsic from intervening absorbers).

\section{What is the True Fraction of Outflows?}

Table~\ref{tab:demographics} summarizes the incidence of outflows in
AGN from several recent surveys (with the redshift and luminosity
ranges of the AGN listed in columns two and three, respectively, and
the ranges in outflow velocity and velocity width listed in columns
four and five, respectively). Inspection of the table shows that the
outflow fraction is dependent both on the characteristics of the
parent sample of AGN used, and on the forms of intrinsic absorption
included. If one only counts BALs observed in higher luminosity AGN
[$\lambda L_\lambda$(3000\,\AA)$\gtrsim 10^{45}$\,erg s$^{-1}$],
then the outflow fraction is 23\,\% \citep{hf03,gan07b}. However,
this is by no means a complete assessment of outflows.

In order to compute a more complete outflow fraction, one must deal
with three issues: (1) cross-talk among classifications, (2)
dependence of frequencies on quasar properties, and (3) mode of the
outflow. Here, we ignore the third issue and focus on the first two.
\citet{cren99} surveyed outflows in lower luminosity [$\lambda
L_\lambda$(3000\,\AA)$\lesssim 10^{45}$\,erg s$^{-1}$] AGN and we
use their result, 59\%, as one benchmark \citep[see also][who finds
a similar percentage based on \ion{O}{6} absorption]{kriss06}. For
higher luminosity AGN, we start with the \citet{hf03} percentage of
23\%, as it is the purest, and most well-defined sample of outflows.
To this, we must add in two things, the contribution from AALs, and
the contribution from high-velocity NALs/mini-BALs. There is general
agreement between \citet{gan01a} and \citet{ves03} that the AAL
fraction is 23--27\,\%. As noted by \citet{gan01a}, quasars that
host broad-absorption lines also tend to host associated absorption.
However, the sample of quasars employed in the \citet{ves03}
estimates explicitly does not include BAL quasars. Thus, while there
is likely some cross-talk between the class of BAL quasars and AAL
quasars, the above range should minimize this effect. Thus, the
outflow fraction counting BALs and AALs (integrated over ``all''
high luminosity AGN) is 46--50\%. For high velocity NALs/mini-BALs,
there is a more sizeable error margin (12--30\%) owing to cross-talk
and dependence on quasar property. Adding this uncertain number
gives our final tally: 57-80\%. This fraction is surprisingly
comparable to that of lower luminosity AGN.

An alternative approach is to begin with the complete sample of
outflows from \citet{gan07b}. Correcting their overall outflow
fraction (2515/5088, 49\%) for quasar selection biases
\citep[following the strategy of][]{dai07}, we find an outflow
fraction of 60$\pm$5\% ($66/287 + 107/287$, see \S\ref{sec:bal}, and
\S\ref{sec:aal}). This falls in the above range, and the only
missing form of outflow from that sample is NALs/mini-BALs at
velocities larger than $\sim30000$\,\kms. We further note that this
explicitly eliminates cross-talk between quasars hosting BALs and
quasars hosting AALs.

\section{Summary}

We conclude that, largely independent of AGN luminosity, 60\% is a good
reference number for the percentage of AGN with intrinsic outflows.  This
number may increase slightly with more thorough searches of parameter
space (e.g., very high velocity outflows).  Until evidence suggests
otherwise, we recommend that quasar outflows be studied as a single phenomenon
whenever possible and that restrictions based on absorber subclass or data
limitations be clearly stated and considered in the interpretation of results.
Catalogs should clearly state their contamination issues and their limitations
for particular applications as appropriate.  A more comprehensive understanding
of the outflow phenomenon awaits us.


\acknowledgements

We wish to thanks the anonymous referee for comments that improved
the quality of the paper. This publication makes use of data
products from the Two Micron All Sky Survey, which is a joint
project of the University of Massachusetts and the Infrared
Processing and Analysis Center/California Institute of Technology,
funded by the National Aeronautics and Space Administration and the
National Science Foundation.  We acknowledge support from the US
National Science Foundation through grant AST 05-07781.


\end{document}

%% file: tab1.tex
\begin{deluxetable*}{llr@{~to~}lr@{\,to\,}lll}
\tablewidth{0pc}
\tabletypesize{\footnotesize}

\tablecaption{Demographics of Outflows}

\tablehead {
& \multicolumn{6}{c}{Ranges} \\
& \multicolumn{6}{c}{\hrulefill} \\
\colhead{Study} &
\colhead{Redshift} &
\multicolumn{2}{c}{$\log \lambda L_\lambda$(3000\,\AA)} &
\multicolumn{2}{c}{Velocity} &
\colhead{Width} &
\colhead{Fraction}\\
& &
\multicolumn{2}{c}{($\log$\,[erg s$^{-1}$])} &
\multicolumn{2}{c}{($10^3$\kms)} &
\colhead{(\kms)}}
\startdata
\citet{cren99}    & $\leq 0.08$ & 42   & 44.8                 &   0  & +2  & $\lesssim 2000$ & 50--70\% \\ \\[-7pt]
\citet{gan01a}    & $\leq 1$    & 44.5 & 46.7                 & $-1$ & +5  & $\lesssim 500$  & 25\% \\ \\[-7pt]
\citet{lb02}      & $\leq 0.5$  & 43.9 & 46.4                 &   0  & +31 & \nodata         & 50\% \\ \\[-7pt]
\citet{hf03}      & 1.5--3.0    & 45.8 & 47                   &  +5  & +25 & $\geq    2000$  & 15\% $\rightarrow$ 23\% \\ \\[-7pt]
\citet{ves03}     & 1.5-3.6     & 45.5 & 47                   &   0  & +5  & $\lesssim 500$  & 27\%  \\
                  &             & \multicolumn{2}{c}{~}       &  +5  & +21 &                 & 18\%  \\ \\[-7pt]
\citet{misawa07}  & 2--4        & 46.5 & 47.6                 &   0  & +5  & $\lesssim 500$  & 23\%  \\
                  &             & \multicolumn{2}{c}{~}       &  +5  & +50 &                 & 30\%  \\ \\[-7pt]
Rodriguez Hidalgo & 1.8-3.5     & \multicolumn{2}{c}{\nodata} &   0  & +5  & 800--3000       & 2\%   \\ \\[-7pt]
et al. (2007)     &             & \multicolumn{2}{c}{~}       &  +5  & +10 &                 & 3\% \\ \\[-7pt]
                  &             & \multicolumn{2}{c}{~}       &  +10 & +25 &                 & 6.7\% \\ \\[-7pt]
                  &             & \multicolumn{2}{c}{~}       &  +25 & +50 &                 & 2.3\% \\ \\[-7pt]
\citet{gan07b}    & 1.7--2.0    & 44.8 & 46.6                 & $-1$ & +40 & $\gtrsim  500$  & 12\% $\rightarrow$ 23\% \\ \\[-7pt]
                  &             & \multicolumn{2}{c}{~}       & $-1$ &  +5 & $\lesssim 500$  & 37\% $\rightarrow$ 37\% \\ \\[-7pt]
\citet{dai07}     & 1.7--4.0    &      &                      &   0  & +25 & $\gtrsim 1000$  & 26\% $\rightarrow$ 40\%
\enddata
\tablecomments{The percentages in the last column indicates the fraction of AGN (in the
redshift and luminosity ranges listed in columns two and three, respectively) that host
intrinsic absorption (with velocities and velocity widths listed in columns four and five,
respectively). An arrow indicates a percentage that has been corrected for possible
selection biases. The luminosities in column three were computed assuming a $h = 0.7$,
$\Omega_\mathrm{m} = 0.3$, $\Omega_\Lambda = 0.7$, $q = 0.5$\ cosmology.}
\label{tab:demographics}
\end{deluxetable*}